\documentstyle[prl,aps,multicol]{revtex}
\input{epsf}
\begin{document}
\draft
\title{Emergence of Fermi-Dirac Thermalization in the Quantum Computer Core}  

\author{Giuliano Benenti$^{(a,b,c)}$, Giulio Casati$^{(b,c,d)}$, and 
Dima L. Shepelyansky$^{(a)}$}

\address{$^{(a)}$Laboratoire de Physique Quantique, UMR 5626 du CNRS, 
Universit\'e Paul Sabatier, F-31062 Toulouse Cedex 4, France}
\address{$^{(b)}$International Center for the Study of Dynamical 
Systems, \\ 
Universit\`a degli Studi dell'Insubria, via Valleggio 11, 22100 Como, Italy}  
\address{$^{(c)}$Istituto Nazionale di Fisica della Materia, 
Unit\`a di Milano, via Celoria 16, 20133 Milano, Italy} 
\address{$^{(d)}$Istituto Nazionale di Fisica Nucleare, 
Sezione di Milano, via Celoria 16, 20133 Milano, Italy} 

%\date{\today}
\date{September 21, 2000}

\maketitle

\begin{abstract}
We model an isolated quantum computer as a two-dimensional lattice 
of qubits (spin halves)    
with fluctuations in individual qubit energies and residual 
short-range inter-qubit couplings.  
In the limit when fluctuations and couplings
are small compared to the one-qubit energy spacing, the  
spectrum has a band structure and we study the quantum computer core
(central band) with the highest density of states. 
Above a critical inter-qubit coupling strength, quantum chaos 
sets in, leading to quantum ergodicity of  eigenstates
in an isolated quantum computer. 
The onset of chaos results in the interaction
induced dynamical thermalization and the occupation numbers well described by
the Fermi-Dirac distribution.  
This thermalization destroys the 
noninteracting qubit structure and sets serious requirements for
the quantum computer operability.    
\end{abstract}
\pacs{PACS numbers: 03.67.Lx, 05.45.Mt, 24.10.Cn}

\begin{multicols}{2}

\narrowtext

\section{Introduction}
\label{intro}

The key ingredient of a quantum computer \cite {steane,springer} is that it 
can simultaneously follow all of the computation paths corresponding 
to the distinct classical inputs and produce a final state which depends 
on the interference of these paths.  
As a result, some computational tasks can be performed much more efficiently  
than on a classical computer. Shor \cite{shor} constructed a quantum algorithm 
which performs large number factorization into prime factors exponentially 
faster than any known classical algorithm. It was also shown by Grover 
\cite{grover} that the search of an item in an unstructured list can be 
done with a square root speedup over any classical algorithm. 
These results motivated a great body of experimental proposals for a 
construction of a realistic quantum computer (see \cite{steane,springer} and 
references therein). At present, quantum gates were realized with cold 
ions \cite{2qubits} and the Grover algorithm was performed for three qubits 
made from nuclear spins in a molecule \cite{3qubits}. 
For a proper operability, it is essential for the quantum computer to remain 
coherent during the computation process. Hence, a serious obstacle to its 
physical realization is the quantum decoherence due to the coupling with the 
external world \cite{CZ,paz,zurek}. 
In spite of that, in certain physical proposals, for example nuclear
spins in two-dimensional semiconductor structures, the decoherence  
time can be many orders of magnitude larger than the time required 
for the gate operations (see for example Refs.\cite{divincenzo,vagner}). 
As a result, one can analyze the operation of an isolated quantum computer 
decoupled from the external world. 

However, even if the quantum computer is 
isolated from the external world and the decoherence time is 
infinite, a proper operability of the computer is not guaranteed.  
As a matter of fact, one has to face a many-body problem for a 
system of $n$ interacting qubits (two level systems): any computer 
operation --a unitary transformation in the Hilbert space of size 
$N_H=2^n$-- can be decomposed into two-qubit gates such as 
controlled-NOT and single qubit rotations \cite{steane,springer}. 
Due to the unavoidable presence of imperfections, the spacing
between the two states of each qubit fluctuates in some  
detuning interval $\delta$. Also, some residual interaction 
$J$ between qubits necessarily remains when the two-qubit 
coupling is used to operate the gates. 

In \cite{GS1,GS2,dima} an isolated quantum computer was modeled 
as a qubit lattice with fluctuations in individual qubit energies 
and residual short-range inter-qubit couplings.  
Similarly to previous studies of interacting many-body systems 
such as nuclei, complex atoms, quantum dots, and quantum spin glasses
\cite{aberg,zelevinsky,flambaum,sushkov,jacquod,pichard,georgeot}, 
the interaction leads to quantum chaos characterized
by ergodicity of the eigenstates and level spacing statistics 
as in  Random Matrix Theory \cite{guhr}.  
The transition to chaos takes place when the interaction strength 
is of the order of the energy spacing between directly coupled 
states \cite{aberg,sushkov,jacquod,pichard,georgeot}. 
This border is exponentially larger than the energy level spacing 
in a quantum computer.  

In this paper we show that the onset of chaos leads to occupation 
number statistics given by the Fermi-Dirac 
distribution. This means that a strong enough interaction plays 
the role of a heat bath, thus leading to dynamical thermalization
for an isolated system.
In such a regime, a quantum computer eigenstate is composed by 
an exponentially large (with $n$) number of noninteracting multi-qubit 
states representing the quantum register states. As a result,   
exponentially many states of the computation basis are mixed 
after a chaotic time scale and the computer operability is destroyed.   
We note that the dynamical thermalization has been discussed in 
other many-body interacting systems in 
\cite{zelevinsky,flambaum,jacquod,borgonovi}. 

The paper is composed as follows. In Section \ref{model}  
we describe our qubit lattice quantum computer model 
\cite{GS1,GS2,dima}; in Section \ref{statistics} we discuss the statistical 
properties of the eigenvalues of this model; in Section \ref{thermalization} 
we study the occupation number distribution and compare different 
definitions for the effective temperature of the system; the conclusions 
are presented in Section \ref{conc}. 

\section{The model}
\label{model} 

We consider a model of $n$ qubits on a two-dimensional lattice 
with nearest neighbors inter-qubit coupling \cite{2D}. The Hamiltonian 
of this model, introduced in \cite{GS1}, reads:  
\begin{equation}
\label{hamil}
H = \sum_{i} \Gamma_i \sigma_{i}^z + \sum_{i<j} J_{ij} 
\sigma_{i}^x \sigma_{j}^x,
\end{equation}
where the $\sigma_{i}$ are the Pauli matrices for the qubit $i$ and the second
sum runs
over nearest-neighbor qubit pairs on a two-dimensional lattice
with periodic boundary conditions applied.
The energy spacing between the two states of a qubit is determined by  
$\Gamma_i=\Delta_0+\delta_i$, with $\delta_i$ randomly and uniformly 
distributed in the interval $[-\delta /2,\delta /2]$. Therefore the 
detuning parameter $\delta$ gives the width of the $\Gamma_i$ distribution 
around its average value $\Delta_0$.  
For generality we choose the couplings $J_{ij}$, which represent the 
residual interaction, randomly and uniformly distributed in the 
interval $[-J,J]$. 
The model (\ref{hamil}) can be considered as a standard generic quantum 
computer model, in which the unavoidable system imperfections generate a 
residual inter-qubit coupling and energy fluctuations.  
In a sense (\ref{hamil}) describes the quantum computer hardware, 
while to study the gate operations in time one should include additional 
time-dependent terms in the Hamiltonian. 
At $J=0$, the noninteracting eigenstates of the model can be written as 
$|\psi_k\rangle=|\alpha_1,...,\alpha_n\rangle$, where 
$\alpha_i=0,1$ marks the polarization of each individual qubit.  
These are the ideal multi-qubit eigenstates of a quantum
computer, the quantum register states used for computer operations.  
For $J \neq 0$, these states are no longer eigenstates of the
Hamiltonian, and the new multi-qubit eigenstates are now linear combinations 
of different quantum register states.  

Here we focus on the case $\delta \ll \Delta_0$, which corresponds
to the situation where fluctuations induced by imperfections are relatively
weak \cite{GS2}. In this case, the unperturbed energy spectrum of (\ref{hamil}) 
(corresponding to $J=0$) is composed of $n+1$ well separated bands, with 
interband spacing $2\Delta_0$.  
Since the $\delta_i$ randomly fluctuate in an interval of size $\delta$,  
each band at $J=0$ (except the extreme ones) has a Gaussian shape of  
width $\approx \sqrt{n} \delta $ \cite{bandwidth}. 
The average number of states inside a band $N_B$ is of the order of 
$N_H/n=2^n/n$, so that the 
energy spacing between adjacent multi-qubit states inside one band is exponentially 
small: $\Delta_n \sim n^{3/2} 2^{-n} \delta$. 
 
In the presence of a residual interaction $J \sim \delta$, the spectrum
still has the above band structure with an exponentially large density of states.
For $J,\delta \ll \Delta_0$, the interband coupling is
very weak and can be neglected.  In this situation, the Hamiltonian 
(\ref{hamil}) is, to a good approximation, described by the 
renormalized Hamiltonian 
\begin{equation}  
\label{hren}
H_{P}=\sum_{k=1}^{n+1} \hat{P_k} H \hat{P_k}, 
\end{equation} 
where $\hat{P_k}$ is the projector on the $k^{th}$ band, so that qubits 
are coupled only inside one band. We concentrate our studies on the central 
band. For an even $n$ this band is centered exactly at $E=0$,
while for odd $n$ there are two bands centered at $E=\pm \Delta_0$,
and we consider the one at $E=-\Delta_0$. The central band corresponds
to the highest density of states, and in a sense represents 
the quantum computer core: an exponentially large number of states 
allows to take advantage of quantum parallelism in 
computer algorithms \cite{steane,springer,shor,grover}.  
On the other hand, quantum chaos and ergodicity first appear 
in this band, which therefore 
sets the limit for operability of the quantum computer. Inside this
band, the system is described by the renormalized Hamiltonian $H_P$ 
which depends only on the number of qubits $n$ and the dimensionless
coupling $J/\delta$. 

\section{Spectral statistics}
\label{statistics} 

The results of Refs.\cite{GS1,GS2} showed that the quantum chaos border 
in (\ref{hamil}) corresponds to a critical interaction $J_c$ 
given by:  
\begin{equation}
\label{Jc}
J_c \approx \frac{C\delta}{n},
\end{equation}
where $C$ is some numerical constant. 
This border is exponentially larger than the energy spacing between 
multi-qubit states $\Delta_n$.
This is in agreement with previous studies of complex interacting 
many-body systems \cite{aberg,sushkov,jacquod,pichard,georgeot}, in which 
the transition to quantum chaos takes place when the interaction 
matrix elements between directly coupled states become larger 
than their energy spacing.   
Since the interaction is of a two-body nature, each noninteracting 
multi-qubit state $|\psi_k\rangle$ has nonzero coupling matrix elements 
only with about $n$ other multi-qubit states. 
Therefore, the number of directly coupled states is much smaller than 
the number of multi-qubit states 
inside the central band, $N_B=n!/([n/2]!(n-[n/2])!)$ 
(we consider the band with the number of spins up given by 
the integer part of $n/2$).  
These couplings induce transitions in an energy interval of 
order $\delta$ (we assume that $J$ is of the order of or smaller than 
$\delta$). 
Therefore the energy spacing between directly coupled 
states is $\Delta_c \sim \delta/n$. The transition to chaos takes 
place for $J=J_c\approx\Delta_c$, which leads to the relation (\ref{Jc}).   

\begin{figure} 
\centerline{\epsfxsize=8cm\epsffile{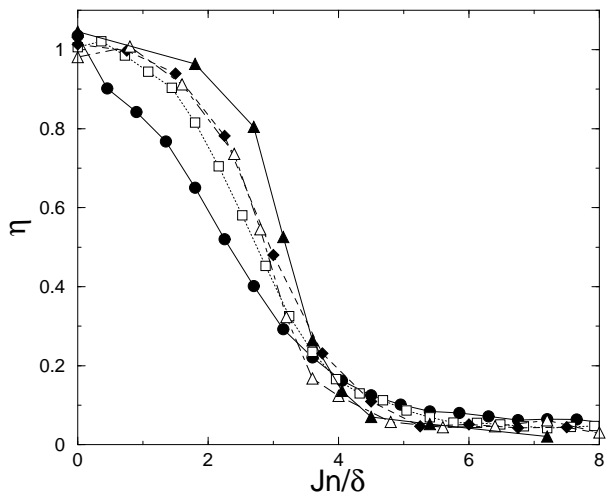}}
\caption{Dependence of $\eta$ on the scaled coupling $Jn/\delta$,   
for $n=9$ qubits (circles, $N_D=10^4$ 
random realizations of $\delta_i,J_{ij}$), $n=12$ (squares, $N_D=10^3$),  
$n=15$ (diamonds, $N_D=45$), $n=16$ (empty triangles, $N_D=23$),  
and $n=18$ (filled triangles, $N_D=3$).The parameter $\eta$ is computed   
for $\pm 5\%$ of states around the center of the energy band.}    
\label{fig1} 
\end{figure} 

The transition to quantum chaos and ergodic eigenstates can be detected 
in the change of the spectral statistics of the system. 
A convenient way is to look at the level spacing statistics $P(s)$, 
which gives the probability to find two adjacent levels whose spacing, 
normalized to the average level spacing, is in $[s,s+ds]$. 
In fact, $P(s)$ goes from the Poisson distribution $P_P(s)=\exp(-s)$ for 
nonergodic states to the Wigner-Dyson distribution 
$P_W(s)=(\pi s/2)\exp(-\pi s^2/4)$, corresponding to Random 
Matrix Theory, for ergodic states \cite{guhr}.    
To analyze the change of $P(s)$ with the coupling $J$ one 
can conveniently use the parameter 
$\eta=\int_0^{s_0}(P(s)-P_W(s))ds/\int_0^{s_0}(P_P(s)-P_W(s))ds$ 
\cite{jacquod}, where $s_0=0.4729...$ is the first intersection 
point of $P_P(s)$ and $P_W(s)$. In this way $P_P(s)$ corresponds to 
$\eta=1$ and $P_W(s)$ to $\eta=0$.  
Fig.\ref{fig1} gives the dependence of the parameter $\eta$ on  
the scaled coupling $Jn/\delta$ at different system sizes, for states 
near the middle of the 
energy spectrum ($\pm 5\%$ of levels around the band center). 
In order to reduce statistical fluctuations, we use 
$3\leq N_D \leq 10^4$ random realizations of $\delta_i,J_{ij}$. 
In this way the total number of spacings $N_S$ is varied in the  
interval $1.4\times 10^4 \leq N_S \leq 1.2 \times 10^5$.   
The Poisson to Wigner-Dyson crossover becomes sharper when 
$n$ increases, suggesting a sharp transition in the thermodynamic
limit. 
We note that, since the number of random $\delta_i,J_{ij}$ values 
is not large, significant fluctuations are present 
in the $\eta$ curves when one changes the random realization. 
For example, for $n=18$, at $J=0.2\delta$, one has 
$\eta=0.44,0.29,0.06$ in the $N_D=3$ random realizations considered. 
This reflects the general property of fluctuations which become 
stronger near the critical transition point. 
On the other hand, even if the number of considered spacings is always 
large, for $n=15,16$ and especially for $n=18$, we have only a small number 
of random realizations, due to the very slow convergence 
of the Lanczos algorithm \cite{cullum} near the band center, where 
the density of states becomes exponentially large. 
These fluctuations 
prevent us from precisely evaluating the critical scaled coupling for 
the Poisson to Wigner transition.  
The minimum spreading of curves is for $\eta(J_c)\approx 0.2$, 
corresponding to $J_cn/\delta\approx 3.7$, in agreement with the 
results of Ref. \cite{GS2}. 
We stress that the chaos border is exponentially larger than the 
multi-qubit level spacing, e.g., for $n=18$, $J_c\approx 0.2\delta \gg 
\Delta_n\approx 7\times 10^{-5}\delta$. 

\begin{figure} 
\centerline{\epsfxsize=8cm\epsffile{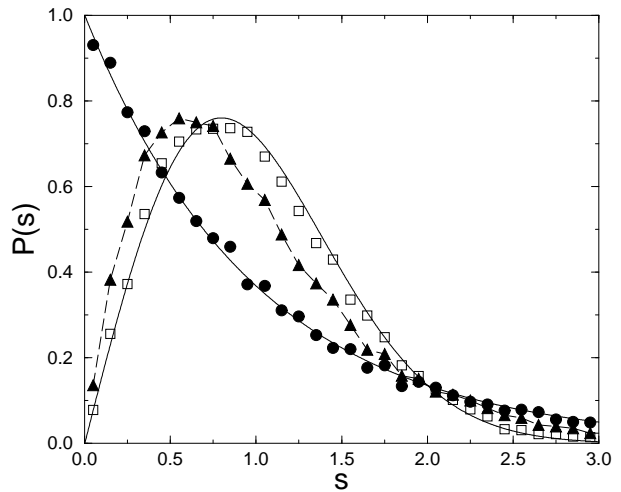}}
\caption{Level spacing statistics corresponding to Fig.\ref{fig1} 
at $n=16$, $J=0.05\delta$ (circles, $\eta=1.01$), $J=0.2\delta$ (triangles, 
$\eta=0.32$), and $J=0.4\delta$ (squares, $\eta=0.05$). Full curves show 
Poisson and Wigner-Dyson distributions.}
\label{fig2} 
\end{figure} 

\begin{figure} 
\centerline{\epsfxsize=8cm\epsffile{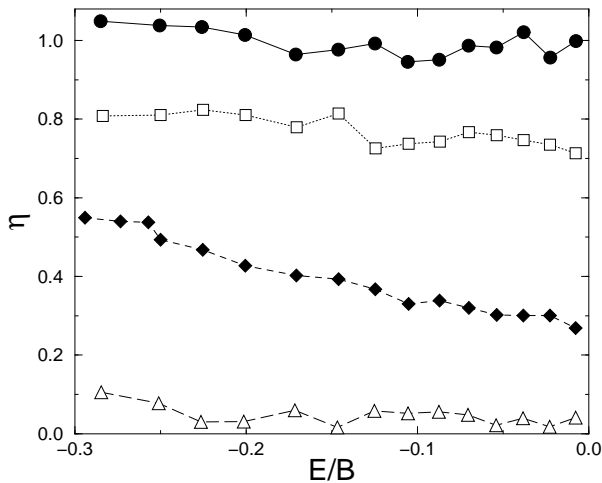}}
\caption{Dependence of $\eta$ on the energy $E$ (normalized to 
the total band width $B$ taken from numerical data) for $n=16$, 
$J=0.05\delta$ (circles), 
$J=0.15\delta$ (squares), $J=0.2\delta$ (diamonds), and $J=0.4\delta$ (triangles). 
Data are averaged over $N_D=15$ random realizations.}
\label{fig3} 
\end{figure}

The level spacing statistics near the band center is shown in 
Fig.\ref{fig2} at different coupling strengths $J$ for $n=16$. 
The transition from the Poisson to 
the Wigner-Dyson statistics is evident. 
In Fig.\ref{fig3} we show that, even away from the band center, the 
transition happens at approximately 
the same critical coupling. 
This is due to the fact that, even if the $n$-body density 
grows exponentially with the excitation energy, the density of 
coupled states remains roughly the same, except near the band edges
\cite{edge}.  
We note that the data are shown only for half of the central band
($E<0$), since the density of states is symmetric around $E=0$ due to the 
presence of an upper bound in the single qubit energies. 

\section{Dynamical thermalization}
\label{thermalization}

The transition in the level spacing statistics reflects a qualitative
change in the structure of the eigenstates \cite{GS1,GS2}.
While for $J \ll J_c$ the eigenstates are very close to 
the quantum register states, for $J>J_c$ each eigenstate 
$|\phi_m\rangle$ becomes a superposition of an exponentially large
number of noninteracting eigenstates $|\psi_k\rangle$. 
It is convenient to characterize the complexity of an eigenstate 
$|\phi_m\rangle$ by the quantum eigenstate entropy 
\begin{equation} 
S_q = -\sum_{k=1}^{N_B} W_{km} \log_2 W_{km}, 
\end{equation} 
where $W_{km}$ is the probability
to find the quantum register state $|\psi_k\rangle$ in the eigenstate 
$|\phi_m\rangle$ of the Hamiltonian 
($W_{km}=|\langle\psi_k|\phi_m\rangle|^2$).  
In this way $S_q=0$
if $|\phi_m\rangle$ is a quantum register state ($J=0$), 
$S_q=1$ if $|\phi_m\rangle$ is equally
composed of two $|\psi_i\rangle$, and the maximum value is 
$S_q=\log_2 N_B$ if all
states equally contribute to $|\phi_m\rangle$.  

Above the chaos border ($J>J_c$) one eigenstate is composed of order 
$2^{S_q}$ quantum register states, mixed inside the Breit-Wigner 
width $\Gamma\sim J^2/\Delta_c\sim J^2 n/\delta$ \cite{GS2}. 
As a result, the residual interaction disintegrates a quantum register 
state over an exponentially large number of states after a chaotic 
time scale $\tau_\chi\approx1/\Gamma$ \cite{GS2,flambaum2}. After this time the quantum 
computer operability is certainly destroyed, unless one can apply  
quantum error-correcting codes (see \cite{steane,springer} and references therein) 
operating on a shorter time scale.  
This destruction takes place in an isolated system, without any 
external decoherence process. It happens due to inter-qubit  
coupling, which can mimic the effect of a coupling with the external 
world. In the following we show that in the quantum chaos regime 
a statistical description of our isolated $n$-qubit system is 
indeed possible, similarly to results found for other  
physical systems in \cite{zelevinsky,flambaum,borgonovi}. 

We concentrate on the distribution of the occupation numbers $n_i$, 
defined as the probability that the qubit (spin) at the site $i$ 
is in its up polarization state. Given an eigenfunction $|\phi_m\rangle$ 
with eigenvalue $E_m$, one can write: 
\begin{equation} 
n_i(m)=\sum_{k=1}^{N_B} W_{km} \langle \psi_k | \hat{n}_i | 
\psi_k \rangle, 
\end{equation} 
where $\hat{n}_i$ is the occupation number operator, and the term 
$\langle \psi_k | \hat{n}_i | \psi_k \rangle$ equals $1$ or $0$ 
depending on whether the spin at the site $i$ is up or down. 

For noninteracting qubits one can write, e.g. for the central band,   
$$
\sum_{i=1}^n n_i(m)=\left[\frac{n}{2}\right],    
$$
\begin{equation} 
\label{FD} 
\sum_{i=1}^n n_i(m) \delta_i=E_m^{'}, 
\quad 
(n_i(m)=0,1), 
\end{equation} 
where $E_m^{'}=E_m/2+\sum_i\delta_i/2$   
($E_m=\sum_i(2n_i(k)-1)\delta_i$). As $n_i(m)=0,1$, the relations 
(\ref{FD}) are the usual ones used to derive the Fermi-Dirac distribution 
for an ideal gas of many noninteracting particles in contact with a 
thermostat. However, here we consider an isolated system of relatively 
few interacting particles. Nevertheless, recent studies 
\cite{zelevinsky,flambaum,borgonovi} have demonstrated that interaction 
can play the role of a heat bath, thus allowing one to use a statistical 
description even in an isolated system with few particles.  
The Fermi-Dirac statistics appears due to the fact that the number of 
spins up/down is fixed and in this way they become equivalent, for 
the purposes of a statistical description, to electrons/holes. 

\begin{figure} 
\centerline{\epsfxsize=8cm\epsffile{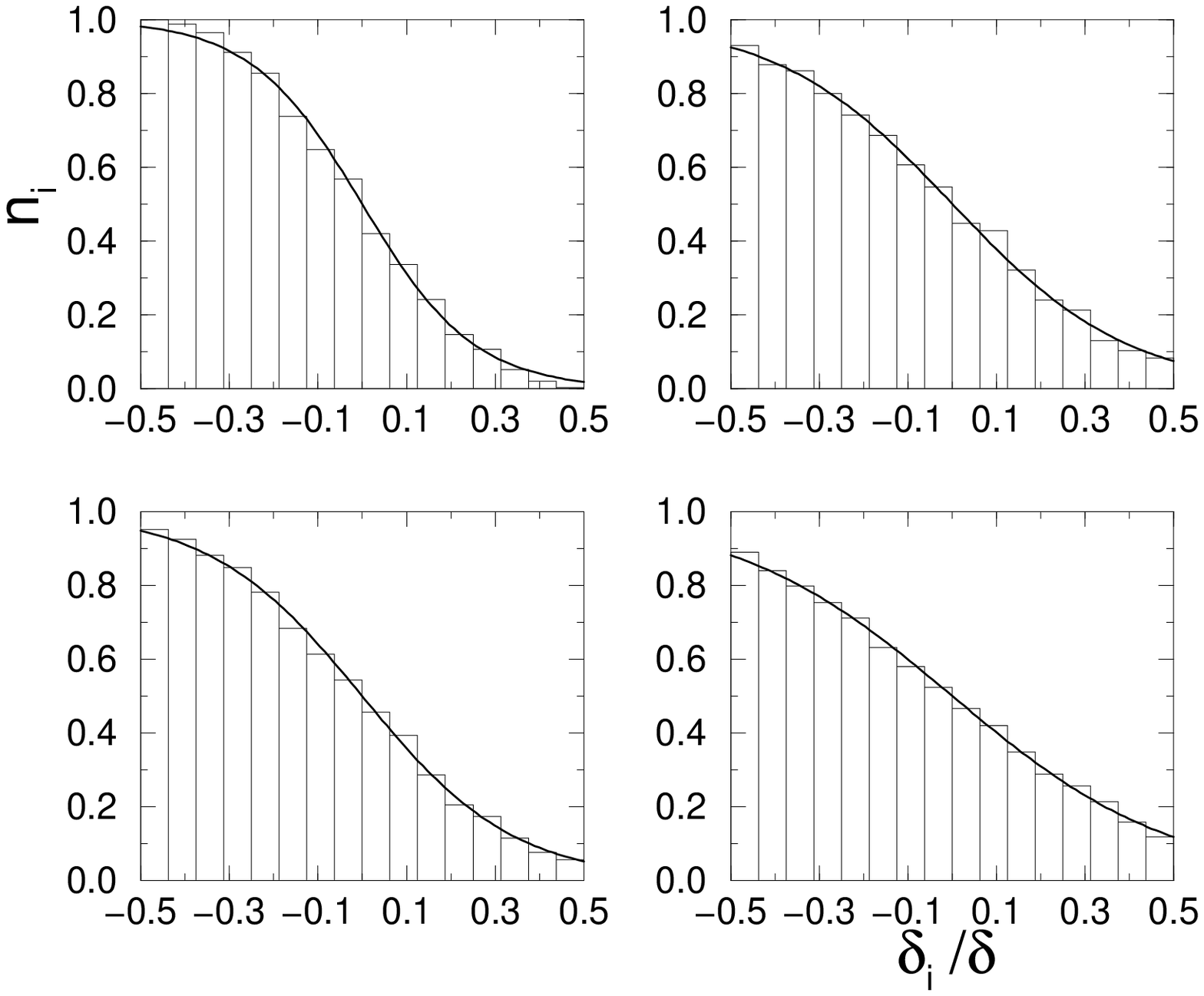}}
\caption{Distribution of the occupation numbers $n_i$ as a function 
of the qubit detunings $\delta_i$, for $n=16$ qubits, coupling 
strength $J=0.03\delta$ (upper figures) and $J=0.3\delta$ (lower figures), 
levels $m=5-10$ (left) and $m=95-100$ (right). Solid lines 
give Fermi-Dirac fits with effective temperature $T_{FD}$. 
Top left: $T_{FD}=0.13\delta$, 
mean excitation energy $\delta E=0.21\delta$, mean quantum eigenstate entropy 
$S_q=0.56$;
top right: $T_{FD}=0.20\delta$, $\delta E=1.14\delta$, $S_q=0.81$; 
bottom left: $T_{FD}=0.17\delta$, $\delta E=0.24\delta$, $S_q=6.03$; 
bottom right: $T_{FD}=0.25\delta$, $\delta E=1.32\delta$, $S_q=8.04$. 
Data are averaged over $N_D=100$ random realizations.} 
\label{fig4} 
\end{figure}

In Fig.\ref{fig4} we show the occupation number distribution, 
averaged over a few consecutive levels $|\phi_m\rangle$ and over $N_D=100$ 
random realization of $\delta_i,J_{ij}$, for $n=16$ qubits, both 
in the integrable regime (top figures) and in the quantum chaos 
regime (bottom figures). We see that in both cases this averaged 
distribution is in very good agreement with the Fermi-Dirac distribution 
\begin{equation} 
\label{nocc} 
n_i^{FD}=
\frac{1}{\exp(\beta(\delta_i+\delta/2-\mu))+1},  
\end{equation} 
where $\mu$ is the chemical potential and $\beta=1/T_{FD}$ is  
the inverse temperature (we set $k_B=1$). Taking into account the 
constraint set by the fixed number of spins up  
($\sum_i n_i^{FD}=[n/2]$), $T_{FD}$ is the only 
fitting parameter. 

The goodness of the fit given by the expression (\ref{nocc}) 
is hardly surprising as statistical distributions are 
obtained for noninteracting particles after a correct counting 
of states. In this procedure, a weak interaction gives a slight 
increase of the system temperature \cite{flambaum}.  

\begin{figure} 
\centerline{\epsfxsize=8cm\epsffile{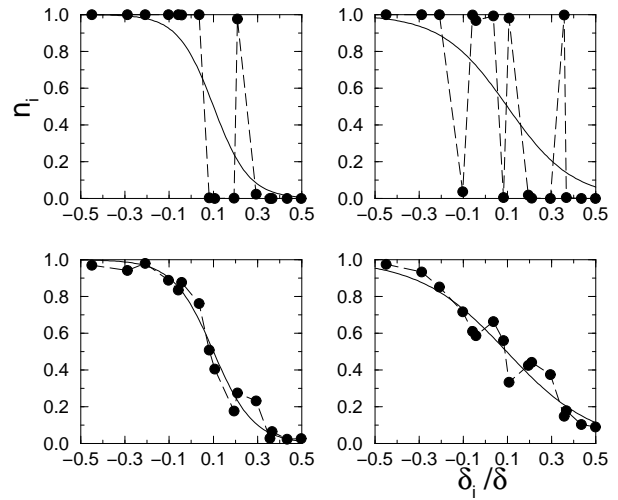}}
\caption{Same as in Fig.\ref{fig4} but for a given random realization
and a single eigenstate for $n=16$ qubits (left: $m=5$; right: $m=100$). 
Top left: $J=0.03\delta$, $T_{FD}=0.08\delta$, 
$\delta E=0.25\delta$, $S_q=0.23$; 
top right: $J=0.03\delta$, $T_{FD}=0.15\delta$, 
$\delta E=0.97\delta$, $S_q=0.49$; 
bottom left: $J=0.3\delta$, $T_{FD}=0.09\delta$, 
$\delta E=0.28\delta$, $S_q=5.85$;  
bottom right: $J=0.3\delta$, $T_{FD}=0.20\delta$, 
$\delta E=1.19\delta$, $S_q=8.41$.}   
\label{fig5} 
\end{figure}

A good agreement between the numerical data in Fig.\ref{fig4} and the 
theoretical distribution (\ref{nocc}) does not mean that 
automatically there is equilibrium and thermalization for a given 
realization. 
This is outlined in Fig.\ref{fig5}, which shows the 
occupation numbers for a single eigenstate of a given random realization. 
In the upper figures 
($J=0.03\delta<<J_c\approx 0.2\delta$) a given eigenstate significantly projects  
only over a single quantum register state ($S_q< 1$) and therefore 
half of the occupation numbers is close to $1$, half close to $0$,    
and the Fermi-Dirac distribution (\ref{nocc}) is very     
far from the actual distribution.   
On the contrary, in the quantum chaos regime (lower figures, 
$J=0.3\delta>J_c$), where a large number of quantum register states are 
mixed in a single eigenstate, there is a good agreement between the occupation 
number distribution and the Fermi-Dirac distribution. 

In order to make quantitative the comparison with the Fermi-Dirac distribution, 
we introduce a parameter which measures the root mean square deviation  
of the actual distribution from (\ref{nocc}): 
\begin{equation} 
\sigma_{FD}(m)=\sqrt{\frac{1}{n}\sum_{i=1}^{n}(n_i(m)-
n_i^{FD}(m))^2. 
}
\end{equation}  
For the case of Fig.\ref{fig5}, we have 
$\sigma_{FD}=6.3\times 10^{-2}$ (top left),  
$\sigma_{FD}=8.8\times 10^{-2}$ (top right) 
and much lower values above the chaos border:  
$\sigma_{FD}=1.4\times 10^{-2}$ (bottom left),   
$\sigma_{FD}=1.7\times 10^{-2}$ (bottom right).  
The maximum value $\sigma_{FD}^{max}=0.5$ is obtained at   
the band center ($T_{FD}=\infty$) for $J=0$, when $n_i=1$ 
for $[n/2]$ spins and $n_i=0$ for the remaining ones.  

We introduce the thermalization border $J_t$ as follows: 
for $J<J_t$ eigenstates close in energy yield completely 
different $n_i$-distributions, for $J>J_t$ the $n_i$-distribution 
is stable with respect to the choice of a specific eigenstate 
in a small energy window. 
The appropriate quantity to be considered in addition to $\sigma_{FD}$ 
is the root mean square 
deviation of the occupation numbers for consecutive eigenstates:  
\begin{equation} 
\sigma_{s}=\sqrt{\frac{1}{n}\sum_{i=1}^{n}(n_i(m+1)-n_i(m))^2}. 
\end{equation}  
For the case of Fig.\ref{fig5}, we have 
$\sigma_s=0.1$ (top left),  
$\sigma_s=0.15$ (top right) 
and much lower values in the chaotic regime:  
$\sigma_s=2.9\times 10^{-2}$ (bottom left),   
$\sigma_s=3.3\times 10^{-2}$ (bottom right).  

\begin{figure} 
\centerline{\epsfxsize=8cm\epsffile{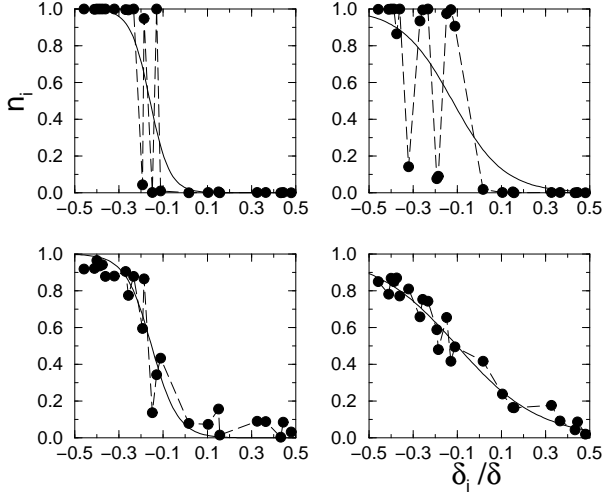}}
\caption{Same as in Fig.\ref{fig5} but for $n=24$ qubits.
Top left: $J=0.05\delta$, $T_{FD}=0.05\delta$, 
$\delta E=0.16\delta$, $S_q=0.51$; 
top right: $J=0.05\delta$, $T_{FD}=0.12\delta$, 
$\delta E=0.64\delta$, $S_q=1.84$; 
bottom left: $J=0.4\delta$, $T_{FD}=0.06\delta$, 
$\delta E=0.38\delta$, $S_q=7.75$; 
bottom right: $J=0.4\delta$, $T_{FD}=0.19\delta$, 
$\delta E=1.16\delta$, $S_q=12.55$.}  
\label{fig6} 
\end{figure}

The conclusions drawn from Fig.\ref{fig5} are also applied to 
Fig.\ref{fig6} where, thanks to the 
effectiveness of the Lanczos algorithm \cite{cullum} near the band 
edges, it was possible to consider a larger number of spins   
($n=24$, corresponding to a very large Hilbert space dimension 
$N_H\approx 1.7\times 10^7$, with $N_B\approx 2.7 \times 10^6$ 
levels in the central band).  

\begin{figure} 
\centerline{\epsfxsize=8cm\epsffile{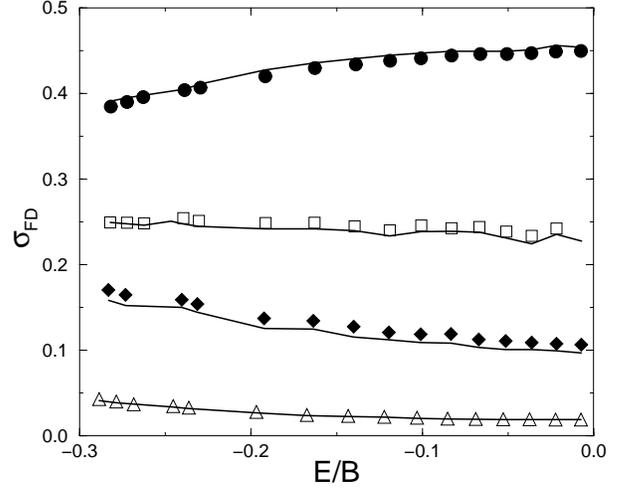}}
\caption{Root mean square deviation $\sigma_{FD}$ of the occupation 
number distribution with respect to the Fermi-Dirac fit as a function 
of the energy $E$ (normalized to the band width $B$), for $n=16$, 
$J=0.05\delta$ (circles), $J=0.15\delta$ (squares), $J=0.2\delta$ (diamonds), and 
$J=0.4\delta$ (triangles). Lines give $\sigma_s/\sqrt{2}$, with $\sigma_s$ 
root mean square deviation of the occupation numbers for consecutive 
levels. Data are averaged over $N_D=2$ random realizations.}
\label{fig7} 
\end{figure}

Fig.\ref{fig7} shows the parameters $\sigma_{FD}$ and $\sigma_s$ 
as a function of the energy, for different values of the coupling strength  
$J$. As for the transition to chaos (Fig.\ref{fig3}), the thermalization 
occurs at approximately the same critical coupling also away from the 
band center. We note that our data give   
$\sigma_s\approx \sqrt{2}\sigma_{FD}$, which can be easily understood as 
follows: $\sum_i(n_i(m+1)-n_i(m))^2\approx\sum_i(n_i(m)-n_i^{FD}(m))^2+     
\sum_i(n_i(m+1)-n_i^{FD}(m+1))^2-      
2\sum_i(n_i(m)-n_i^{FD}(m))(n_i(m+1)-n_i^{FD}(m+1))$, 
the last term in the sum becoming negligible after ensemble 
averaging. 
We stress that the very good agreement between the parameters 
$\sigma_{FD}$ and $\sigma_s$ implies that the Fermi-Dirac distribution 
emerges when the occupation number distribution is statistically stable, 
i.e. the system is thermalized. 
The curve for $J=0.05\delta$ lowers near the band edge since 
for a small excitation energy $\delta E$ a small number 
$n_{eff}\sim (n\delta E/\delta)^{1/2}$ of spins 
(``fermions'') is available for fluctuations in the  
vicinity of the Fermi level (see the note \cite{edge}). 

\begin{figure} 
\centerline{\epsfxsize=8cm\epsffile{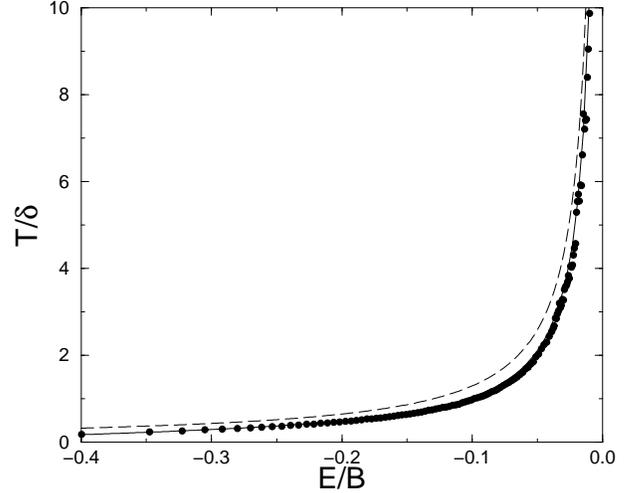}}
\caption{Dependence of different definitions of temperature $T$ 
on the scaled energy $E/B$, for $n=16$, $J=0.3\delta$, $N_D=2$:
$T_{FD}$ (circles), $T_{can}$ (full curve), and $T_{th}$ 
(dashed curve).}
\label{fig8} 
\end{figure}

It is interesting to compare the temperature $T_{FD}$ obtained 
from the Fermi-Dirac fit with different definitions of temperature
\cite{zelevinsky,flambaum,borgonovi}, 
which are known to be equivalent at the thermodynamic limit. 
First of all we use the canonical expression 
\begin{eqnarray} 
E(T_{can})=
\frac{\displaystyle{\sum_{m=1}^{N_B}}E_m
\displaystyle{\exp\left(-\frac{E_m'}{T_{can}}\right)}}{
\displaystyle{\sum_{m=1}^{N_B}}
\displaystyle{\exp\left(-\frac{E_m'}{T_{can}}\right)}}, 
\end{eqnarray}  
where $E_m$ are the exact eigenenergies of the interacting system. 
The very good agreement between $T_{FD}$ and $T_{can}$ 
(see Fig.\ref{fig8}) supports the validity of a statistical  
description for our isolated quantum computer model. 
This means that in such closed system the inter-qubit residual 
interaction plays the role of a heat bath in an open system.  
In particular, we expect that residual interaction mimics  
to a certain extent the effect of coupling to external world.     

Finally we compare the effective temperature $T_{FD}$ with 
the standard thermodynamic temperature $T_{th}$, defined by 
\begin{equation} 
\frac{1}{T_{th}}=\frac{dS_{th}}{dE'}=\frac{d\ln\rho}{dE'},
\end{equation}   
where $S_{th}=\ln\rho$ is the thermodynamic entropy, with 
$\rho$ density of states (here $E'$ replaces $E_m'$). A Gaussian 
fit, $\rho=(1/\sqrt{2\pi}\sigma)\exp(-E'^2/\sigma^2)$, 
provides an excellent 
approximation to the actual data for the density of states, 
with the exception of the first few levels near the band edges. 
Therefore one gets $T_{th}=-\sigma^2/E'$, with $\sigma^2$ variance  
of the Gaussian fit. Fig.\ref{fig8} shows that the difference between 
$T_{th}$ and $T_{FD}\approx T_{can}$ increases as one moves 
away from the band center. 
This is due to the fact that, contrary to the noninteracting 
case (see Eq.(\ref{FD})), the actual system energy $E'$ is different 
from $\sum_i n_i \delta_i$, since the total energy is renormalized 
due to interaction \cite{flambaum}. 
We remark that, not only on average but also for a single eigenstate of
a given realization, the agreement between $T_{FD}$ and $T_{can}$ is good, 
while $T_{th}$ becomes closer to $T_{FD}$ and $T_{can}$ when 
the excitation energy increases.  
In the case of Fig.\ref{fig5} we obtain
$T_{FD}=0.08\delta$, $T_{can}=0.07\delta$, 
$T_{th}=0.19\delta$ (top left),  
$T_{FD}=0.15\delta$, $T_{can}=0.16\delta$, 
$T_{th}=0.24\delta$ (top right),  
$T_{FD}=0.09\delta$, $T_{can}=0.08\delta$, 
$T_{th}=0.22\delta$ (bottom left),  
$T_{FD}=0.20\delta$, $T_{can}=0.19\delta$, 
$T_{th}=0.29\delta$ (bottom right).   
We also note that the effective temperature diverges at the band center and 
is negative in the upper part of the spectrum. This is typical of models 
with an upper bound in the single particle energies, in which the 
density of states has a maximum.  

\begin{figure} 
\centerline{\epsfxsize=8cm\epsffile{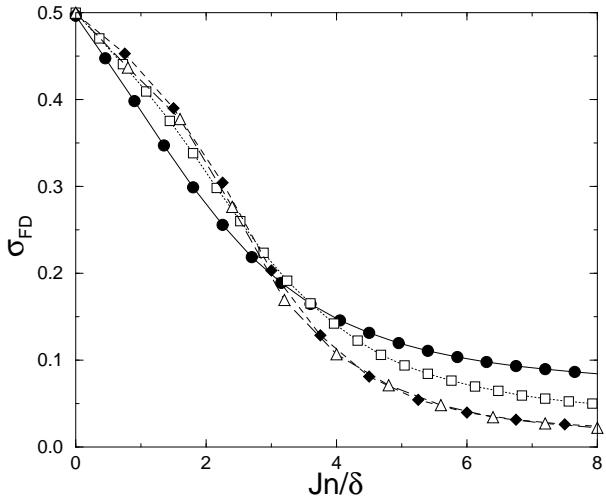}}
\caption{Dependence of $\sigma_{FD}$ on the scaled coupling $Jn/\delta$,
for $n=9$ (circles, $N_D=2\times 10^3$ random 
realizations of $\delta_i,J_{ij}$), $n=12$ (squares, $N_D=2\times 10^2$), 
$n=15$ (diamonds, $N_D=20$), and $n=16$ (triangles, $N_D=10$).
The parameter $\sigma_{FD}$ is computed for $\pm 5\%$ of states 
around the band center. 
Transition to thermalization takes place at $J_t n/\delta\approx 3.2$.}
\label{fig9} 
\end{figure} 

The dependence of the thermalization parameter $\sigma_{FD}$ on the 
scaled coupling $Jn/\delta$ at different system sizes is shown in 
Fig.\ref{fig9} 
(very similar results are obtained for $\sigma_s/\sqrt{2}$). 
Our data show that the crossover to a thermalized distribution sharpens 
when the number of qubits increases, in a way consistent with a sharp 
thermalization border in the thermodynamic limit, at 
$J_t n/\delta\approx 3.2$.          
The similarity between the results of Fig.\ref{fig1} ($J_c n/\delta \approx 3.7$)  
and Fig.\ref{fig9} leads us to the conclusion
that the chaos border coincides with the thermalization 
border. This looks quite natural since the Poisson level spacing statistics   
indicates the existence of uncoupled parts in the whole system, thus 
preventing thermalization. 
On the contrary, in the chaotic regime each eigenfunction spreads  
over an exponentially large number $N\approx 2^{S_q}$ of quantum register 
states, resulting in the Wigner-Dyson statistics. 
In this regime the fluctuations of eigenstate components are 
Gaussian \cite{flambaum} and therefore, according to the central limit 
theorem, the fluctuations of the occupation numbers are small: 
$\Delta n_i \propto N^{-1/2} \ll 1$. For this reason eigenstates 
close in energy give similar $n_i$-distributions, which means that 
there is equilibrium in the statistical sense.       

\section{Conclusions}
\label{conc}

The results presented in this paper show that the residual inter-qubit coupling
can lead to quantum chaos and a statistical description of the 
occupation numbers in close agreement with the Fermi-Dirac distribution. 
We stress that the transition to quantum chaos is an internal process which
happens in a perfectly isolated system with no coupling to the external world.
Nevertheless, the thermalization which appears in this closed system due 
to inter-qubit coupling can mimic the effect of an external thermal bath.    
Above the chaos/thermalization border the quantum register states are 
destroyed after the chaotic time scale and therefore serious restrictions  
are set to the quantum  computer operability.  
However, below this border the imperfection effects only 
slightly disturb the ideal multi-qubit states and the quantum computer 
can operate properly in this regime. 

We thank Bertrand Georgeot for stimulating
discussions, and the IDRIS in Orsay and the CICT in Toulouse for access to 
their supercomputers.

\end{multicols} 


\vskip -0.5cm
\begin{thebibliography}{99}
\bibitem{steane} A. Steane, Rep. Progr. Phys. {\bf 61}, 117 (1998).
\bibitem{springer} {\it The Physics of Quantum Information}, edited 
by D. Bouwmeester, A. Ekert, and A. Zeilinger (Springer-Verlag, 
Berlin, 2000). 
\bibitem{shor} P. Shor, in {\it Proceedings of the $35$-th Annual 
Symposium on Foundations of Computer Science}, edited by 
S. Goldwasser (IEEE Computer Society Press, Los Alamitos, CA, 1994), 
p. 124. 
\bibitem{grover} L.K. Grover, Phys. Rev. Lett. {\bf 79}, 325 (1997);   
{\it ibid.} {\bf 80}, 4329 (1998).  
\bibitem{2qubits} C. Monroe, D.M. Meekhof, B.E. King, W.M. Itano, and 
D.J. Wineland, Phys. Rev. Lett. {\bf 75}, 4714 (1995). 
\bibitem{3qubits} L.M.K. Vandersypen, M. Steffen, M.H. Sherwood, 
C.S. Yannoni, G. Breyta, and I.L. Chuang, Appl. Phys. Lett. 
{\bf 76}, 646 (2000). 
\bibitem{CZ} J.I. Cirac and P. Zoller, 
Phys. Rev. Lett. {\bf 74}, 4091 (1995).  
\bibitem{paz} C. Miquel, J.P. Paz, and R. Perazzo, 
Phys. Rev. A {\bf 54}, 2605 (1996).
\bibitem{zurek} C. Miquel, J.P. Paz, and W.H. Zurek, 
Phys. Rev. Lett. {\bf 78}, 3971 (1997).
\bibitem{divincenzo} D.P. Di Vincenzo, Science {\bf 270}, 255 (1995). 
\bibitem{vagner} D. Mozyrsky, V. Privman, and I.D. Vagner,
cond-mat/0002350.
\bibitem{GS1} B. Georgeot and D.L. Shepelyansky, 
Phys. Rev. E {\bf 62}, 3504 (2000).
\bibitem{GS2} B. Georgeot and D.L. Shepelyansky, 
quant-ph/0005015, Phys. Rev. E (in press). 
\bibitem{dima} D.L. Shepelyansky, quant-ph/0006073. 
\bibitem{aberg} S. {\AA}berg, Phys. Rev. Lett. {\bf 64}, 3119 (1990).
\bibitem{zelevinsky} M. Horoi, V. Zelevinsky, and B.A. Brown, 
Phys. Rev. Lett. {\bf 74}, 5194 (1995);  
V. Zelevinsky, B.A. Brown, N. Frazier, and M. Horoi, 
Phys. Rep. {\bf 276}, 85 (1996). 
\bibitem{flambaum} V.V. Flambaum, F.M. Izrailev, and G.~Casati, 
Phys. Rev. E {\bf 54}, 2136 (1996); V.V. Flambaum and F.M. Izrailev, 
Phys. Rev. E {\bf 56}, 5144 (1997).
\bibitem{sushkov} D.L. Shepelyansky and O.P. Sushkov, Europhys. Lett. 
{\bf 37}, 121 (1997). 
\bibitem{jacquod} Ph. Jacquod and  D.L. Shepelyansky, 
Phys. Rev. Lett. {\bf 79}, 1837 (1997).
\bibitem{pichard} D.Weinmann, J.-L. Pichard, and Y. Imry,
J. Phys. I France {\bf 7}, 1559 (1997).
\bibitem{georgeot} B. Georgeot and D.L. Shepelyansky,
Phys. Rev. Lett. {\bf 81}, 5129 (1998).
\bibitem{guhr} For a review see, e.g., T. Guhr, 
A. M\"uller-Groeling, and H.A. Weidenm\"uller, 
Phys. Rep. {\bf 299}, 189 (1998).
\bibitem{borgonovi} F. Borgonovi, I. Guarneri, F.M. Izrailev, 
and G. Casati, Phys. Lett. A {\bf 247}, 140 (1998). 
\bibitem{2D} We consider a two dimensional lattice for the sake of 
generality. In fact, the corresponding one-dimensional model can be 
mapped by the Wigner-Jordan transformation into a system of noninteracting 
fermions and therefore it is always integrable, see for example 
A.P. Young, Phys. Rev. B {\bf 56}, 11691 (1997).   
\bibitem{bandwidth} The majority of states 
are inside this interval, while, for the bands near to the center, the 
total band width is $B\approx n\delta/2$. This is due to rare events in 
the sum of $n$ random numbers. 
\bibitem{cullum} J. Cullum and R.A. Willoughby, J. Comp. Phys. 
{\bf 44}, 329 (1981). 
\bibitem{edge} At $J=0$ in the ground states the 
$[n/2]$ spins of lowest $\delta_i$ are up, the remaining ones  
are down. Then the low energy excitations 
(magnons) exchange the polarizations of a couple $(i,j)$ of qubits  
with energies $(\delta_i,\delta_j)$ close to the effective Fermi energy 
$\epsilon_F\sim\delta/2$.   
Following standard estimates \cite{jacquod}, the number of 
effectively interacting qubits at a temperature $T\ll\epsilon_F$ is given 
by $n_{eff}\sim Tn/\epsilon_F\sim (n \delta E/\delta)^{1/2}$, with 
$\delta E\sim n_{eff}T$ excitation energy with respect to the 
$n$-qubit ground state. This changes the chaos 
border (\ref{Jc}) to $J_c(\delta E)\sim \delta^{3/2}/\sqrt{n\delta E}$.  
\bibitem{flambaum2} V.V. Flambaum, quant-ph/9911061. 
\end{thebibliography}
\end{document}